\documentstyle[prl,aps,twocolumn]{revtex}
\draft

\title{Creep in One Dimension and Phenomenological
Theory of Glass Dynamics}
\author{Pierre Le Doussal$^{(1,3,*)}$ and Valerii M. Vinokur$^{(2,3)}$}
\address{$^{(1)}$CNRS- Laboratoire de Physique Th\'eorique de l'Ecole
Normale Sup\'erieure, 24 rue Lhomond, F-75231 Paris}
\address{$^{(2)}$Materials Science Division, Argonne National
Laboratory, Argonne IL 60439}
\address{$^{(3)}$ITP, University of California, Santa Barbara CA
93106}

\date{\today}
\begin{document}
\maketitle
\widetext
\begin{abstract}

The dynamics of a glass transition is
discussed in terms of the motion of a particle
in a one dimensional correlated random potential.
An exact calculation of the velocity $V$
under an applied force $f$ demonstrates a variety of
dynamic regimes depending on the range of correlations.
In a gaussian potential with correlator
$C(x) = x^{\gamma}$, we find a transition from
ohmic behaviour ($\gamma <0$) to
creep motion $V \sim \exp(- const/f^{\mu} )$
($0<\gamma <1$). This provides a generic picture
of the glass transition in systems where long range correlations
in the effective disorder develop due to elasticity
such as elastic manifolds subject to quenched disorder
and the vortex glass transition in superconductors.

\end{abstract}
\pacs{74.60.Ge, 05.20.-y}
\narrowtext


The driven dynamics of physical systems which can be modelled as
elastic manifolds in quenched random media has received a lot of attention
recently. Prominent examples of such systems are the roughening of
domain walls \cite{huse_henley}, directed polymer growth
\cite{kardar_parisi_zhang},
motion of dislocations in disordered media
\cite{vinokur_dislocations}, dynamics of charge density waves
\cite{fukuyama_lee_cdw}, and surface growth in a random
environment \cite{parisi_surface}.
This recent interest was partly motivated by extensive
studies of vortex dynamics in high-Tc superconductors. It was shown
that this problem is related to the dynamics of an elastic manifold
subject to quenched disorder
\cite{revue_russes,fisher_vortexglass_long,feigelman_collective,nelson_columnar_long},

and a significant progress in
{\it qualitative} understanding was achieved.
The motion is viewed as a sequence
of thermally activated jumps of the optimal "cell"
of the manifold from one metastable state to the next as
favored by the applied force.
In most of these systems the
activation barriers for such a motion, which we will
refer to as creep motion, depends on the
applied force $f$ and diverges as $f  \rightarrow 0$,
giving rise to a strongly non linear
velocity versus applied force dependence. The unlimited growth
of the creep barriers is taken now as the characteristic
feature and operational definition of glassy dynamics.
Usually \cite{feigelman_collective}
the barriers grow as $U_{B}(f) \sim f^{-\mu}$
leading to a typical
velocity vs applied force dependence (or I-V curve)
of the form $V = e^{-1/(T f^{\mu})}$ in the creep regime.

While there is now a consistent qualitative picture of the low temperature
creep motion \cite{ioffe,revue_russes}, it is based mostly on scaling
arguments
and numerical simulations, and a rigorous analysis is still lacking.
Although interesting new results have been obtained recently
for the non-equilibrium dynamics of mean field models of glasses
\cite{kurchan} and for particle dynamics in an infinite-dimensional
random potential \cite{horner,Lele}, a general analytical derivation in
physical
models remains an unsolved problem.
In the absence of a rigorous analysis of realistic
physical  \newline\vskip 1.27truein\noindent
situations one is seeking for models
which are simple enough to be treated analytically and
yet are able to mimic the large diversity of dynamics of
real glassy systems. A well known example
is the problem of a single particle driven by an external force $f$
and subject to a one dimensional random force field
with Gaussian short range correlations. The term random force
means that the correlator of the random potential $U(x)$
is $\overline{(U(x)- U(y))^2 } \sim \Delta |x-y|$
where $\Delta$ is characterizes
the strength of the random potential. This model is known
as Sinai's model and has long been a subject
of extensive studies
\cite{sinai,solomon,derrida_pomeau,derrida,bouchaud_1d,feigelman_aging_1d,pld_prl,vinokur_dislocations}
The remarkable result obtained for this model is that
even at finite temperature the mobility vanishes below a threshold force
$f_{th} \sim \Delta/2 T$. Moreover this system was
found to exhibit anomalous diffusion, and aging phenomena
\cite{feigelman_aging_1d}
very much like what is observed in spin glasses
\cite{ocio,kurchan}.

In this Letter we study a wider class of one dimensional
random potentials with arbitrary correlations and find that
our model can reproduce most of the existing regimes
of glassy dynamics. The Letter is organized as follows.
First we derive an exact formula for the velocity in
a one dimensional medium. The result is
physically transparent and simple enough to allow for
analytical treatment of very general potentials. Then we study
Gaussian random potentials with a general correlator such that
at large distances $|x-y| \gg a$:
\begin{equation}
\overline{(U(x)- U(y))^2 } \equiv K (x-y) \sim \Delta |x-y|^\gamma
\end{equation}
and in Fourier space
$\tilde{K}(q)=\overline{U(q)U(-q)} \sim_{q \to 0} q^{-(1+\gamma)}$
with
$K(z)=2 \int_{-\Lambda}^{\Lambda} \frac{dq}{2 \pi} (1-\cos q z) \tilde{K}(q)$.
We find that there are several regimes depending on
how correlated the potential is.
If correlations are short range ($\gamma <0$),
we recover a 'viscous flow' (ohmic) regime where
linear response $V \sim f$ holds for small $f$.
If correlations grow with $0< \gamma <1$ we
find a new creep regime $V \sim \exp( -1/f^\mu)$.
The case $\gamma=0$ corresponds to the transition
between these two regimes and we find a power law
critical power law $V$ versus $f$ dependence, reminiscent of the
vortex glass transition behaviour \cite{fisher_vortexglass_long}. Finally
$\gamma=1$ corresponds to Sinai's case where $V=0$ below a threshold
force.

This suggests a deep connection with the motion of
elastic manifold in a random medium (such as vortex systems
in type II superconductors). In these systems creep behaviour
arises from the interplay of the elasticity and pinning potential
\cite{revue_russes} which both determine the creep barriers $U_b$.
The bare pinning potential is uncorrelated but the elasticity of
the manifold generates long range correlations in the effective
potential $U_b$ that determines the motion of the manifold.
The role of elasticity as a tuning
mecanism for correlations becomes transparent upon
noticing that the formation of the vortex glass is caused
by a drastic change in elastic properties.
Namely, the onset of shear modulus at the freezing point
develops long range correlations
in the vortex system \cite{revue_russes}.
A phenomenological approach to describe a generic glass transition
is to introduce a correlation length $\xi_G$ characterizing the
spatial range of critical correlations, which diverges at the
transition. The choice of $K(x)$ on both sides of the transition is
dictated by physical considerations.
In the correlated phase (i.e glass phase)
the natural choice is a Gaussian random potential with
correlator of the form:
\begin{equation} \label{corr1}
K (x-y) \sim
\Delta ( (\frac{|x-y|}{\xi_{G}})^\gamma + \log(\frac{|x-y|}{a} )
\end{equation}
The first term which dominates at
large distance describes long range correlations in the random
potential and generalizes Sinai's model. The
second term describes the behaviour at the critical point $\xi_G=\infty$.
Indeed  one has $\gamma=0$ and
$\tilde{K}(q) \sim 1/q$ at the transition
and thus $K(x) \sim \Delta ~ \ln x$.
The form (\ref{corr1}) is an interpolation
resulting from the crossover between the critical fixed point and
the fixed point describing the glass phase.
In the uncorrelated phase correlations
are short range and one chooses a correlator
as  $\tilde{K}(q)=1/\sqrt{q^2 + (\xi_G)^{-2}}$, i.e
$K(x)=K_0(x/\xi)$,
so as to reproduce the critical behaviour for $a \ll x \ll \xi_G$.
An identical scenario
was demonstrated using RG for the correlations in the free
energy landscape at the glass
transition in surface growth models, such as the directed polymer
in $d \ge 2+1$ \cite{nattermann}

On a mathematical level the present model is the $d=0$, $n=1$ version
of the problem of the dynamics of manifolds of internal dimension $d$, in a
$n$ dimensional space. Remarkably, the case $d=0$ and $n \to \infty$
was recently studied by completely different
techniques and seems to exhibit similar regimes \cite{horner}.

We consider the Langevin diffusion in the one dimensional
quenched random potential $U(x)$ in presence of a global bias $f$
and thermal white noise $\eta(t)$:
\begin{equation} \label{langevin}
\frac{dx(t)}{dt} = - \nabla U(x(t)) + f  + \eta(t)
\end{equation}
with $\langle \eta(t) \eta(t')
\rangle = 2 T \delta(t-t')$ and $T$ is the temperature.
The probability density $P(x,t)$ and the current $J(x,t)$ satisfy:
\begin{equation} \label{fplanck}
\frac{\partial P(x,t)}{\partial t} = - \nabla J(x,t)
\end{equation}
with $J(x,t)= - T \nabla P(x,t) + (f - \nabla U(x)) P(x,t)$.

To derive the analytic expression for the velocity $V$ we
generalize to continuum models the
method introduced by Derrida \cite{derrida} for
discrete hopping problems.
We consider an infinite periodic environment, i.e
a periodic random force $\nabla U(x)$,
of period $L$. The limit $L \rightarrow \infty$ is taken at the end
\cite{footnote1}.
One defines the periodized probability $\tilde{P}(x) = \sum_k P(x+k L)$
which obeys the same equation (\ref{fplanck}) as $P$,
and corresponds to diffusion on a periodic ring of size $L$.
Using (\ref{fplanck}) the velocity for the particule on the infinite line
can be expressed as:
\begin{eqnarray} \label{velo}
\frac{d <x(t)>}{dt}
=&& - \int_{-\infty}^{+\infty}  dx x \nabla J
= \int_{-\infty}^{+\infty} dx J(x,t)
\nonumber \\
&&\
  = \int_{0}^{L} dx \tilde{J}(x,t)
\end{eqnarray}
where $\tilde{J}(x,t)= - T \nabla \tilde{P}(x,t) + (f
- \nabla U(x)) \tilde{P}(x,t)$. At long time $\tilde{J}(x,t)$ goes to a
constant $\tilde{J}$ and the asymptotic velocity $V$ is
exactly given by $V = \tilde{J} L$.
To find $\tilde{J}$ for a fixed $L$ and disorder
configuration one must solve the stationarity equation:
\begin{equation}
T \frac{\partial \tilde{P}(x)}{\partial x}
+ ( \nabla U(x) - f) \tilde{P}(x) = - \tilde{J}
\end{equation}
with the two additional conditions
$\tilde{P}(0)=\tilde{P}(L)$ and $\int _0^L dx \tilde{P}(x) = 1$.
The stationary solution with zero
current $\tilde{J}=0$, i.e the Gibbs distribution
$P_0(x)=\exp(\frac{1}{T}(-U(x)+ f x))$
does not, in general, satisfy the periodic boundary conditions. Thus
$V$ can be found from the solution with non-zero
current:
\begin{eqnarray} \label{pstat}
\tilde{P}(x)=&& \frac{\tilde{J}}{T} \bigg(
\frac{ \int_0^L dy e^{(U(y)- U(x) + f(x-y))/T }}{
1- e^{( U(L)- U(0)- f L)/T}} \nonumber \\
&&\
- \int_0^x dy e^{ ( U(y)- U(x)
+ f(x-y))/T } \bigg)
\end{eqnarray}
$\tilde{J}$ and thus $V$ follow from
the normalization condition for $\tilde{P}$. In the
limit $L \to \infty$, imposing the restriction $U(0)=U(L)$,
unimportant for $f>0$, (\ref{pstat}) simplifies to:
\begin{equation}
\tilde{P}(x)= \frac{\tilde{J}}{T} \int_0^{+\infty} dz ~
e^{( U(x+z)- U(x) - f z)/T}
\end{equation}
and one gets the general formula for $V$:
\begin{equation} \label{velocity}
\frac{1}{V} = \frac{1}{T} \int_{0}^{+\infty}
dz ~ e^{- f z/T} ~ \langle e^{ ( U(x+z) - U(x))/T} \rangle_x
\end{equation}
valid for an arbitrary potential $U(x)$.
$\langle A \rangle_x$ denotes the translational
average
$\langle A \rangle_x =
\lim_{L \rightarrow \infty} L^{-1} \int_0^L dx A(x)$. The average in
(\ref{velocity}) exists quite generally
and is independent of the configuration of the random potential, i.e the
velocity
is self-averaging. The physical interpretation of (\ref{velocity}) in terms
of Arrhenius waiting time is transparent.
The average waiting time $1/V$ is a sum of Boltzman weights associated with
the barriers the particle must overcome to move in
the direction of the driving force. The highest barriers $U(x+z)-U(x)$ with
$z>0$,
produce the largest waiting times.

The expression (\ref{velocity}) reveals immediately
several general features. At large $f$ one has $V \approx f$.
At small force $f \to 0$, the response will be linear only if the barriers
saturate, i.e do not grow at large distance. If the potential is
uncorrelated at
large distances such that
$\langle e^{ (U(x+z) - U(x))/T } \rangle \to
\langle e^{ U/T } \rangle \langle e^{ - U/T } \rangle$
when $z \to \infty$, then:
\begin{equation} \label{diffcoeff}
V \propto \frac{D}{D_0} f = \frac{f}{
\langle e^{ U/T } \rangle \langle e^{ - U/T } \rangle }
\end{equation}
where $D$ and $D_0$ are the diffusion coefficients in presence and
in absence of disorder, respectively and the Einstein relation holds.
When $D \ll D_0$ the $V$-$f$ curve
will show strong nonlinearity at intermediate scales where the
transition between low force and high force regime
of motion occurs.

We now turn to a detailed study of gaussian disorder with correlator $K(x)$.
Upon averaging over disorder (\ref{velocity} yields:
\begin{equation} \label{disorder}
\frac{1}{V}  =  \frac{1}{T} \int_0^\infty dx
\exp(- \frac{f x}{T} + \frac{K(x)}{2 T^2})
\end{equation}
The choice of $K(x)$ as in (2) gives rise to several regimes
of particle dynamics depending on the range of the correlations of the random
potential.

{\it Sinai's case $\gamma=1$}: For $\gamma > 1$ the integral in
(\ref{disorder}) {\it diverges} and the velocity is zero. Sinai's model
corresponds to $\gamma = 1$ and appears as a marginal case
where the integral (\ref{disorder}) diverges for $f < f_{th}= \Delta/(2 T
\xi_G)$
and $V=0$ while $V=f-f_{th}$ exactly for $f>  f_{th}$, in agreement with
previous results
\cite{sinai,solomon,derrida_pomeau,bouchaud_1d,feigelman_aging_1d}.
The system with $\gamma =1$ exhibits algebraic distributions of waiting times
which gives rise to aging phenomena [20]. Therefore Sinai's model mimics
essential aspects of the spin-glass behavior.
Interestingly, this $V$ versus $f$ dependence mimics also the dry friction
phenomenon.

{\it Creep motion $0<\gamma<1$}:
In the intermediate case $0< \gamma <1$ one finds the ''creep'' dynamics
regime. Defining the dynamical exponent $z = 2 + (\Delta/2 T^2)$
and the characteristic
force $f_c=(\Delta/2 T^2)^{1/\gamma}~T/\xi_G$ one arrives at:
\begin{eqnarray} \label{H}
\frac{T}{V} =&&  \frac{1}{{f_c}^{z-1}} ~~ H(f/f_c) \nonumber \\
&&\
\qquad  H(y)=\int_0^\infty dv ~ v^{z-2} ~ \exp( - y v + v^\gamma )
\end{eqnarray}
Note that  $f_c$ reduces to the threshold force $f_{th}$
when $\gamma \to 1$.
At $\gamma < 1$ the sharp threshold disappears but
at $f \ll f_c$ the $V$ versus $f$ dependence shows
strongly nonlinear behaviour with
an essential singularity at small $f$. Using the steepest descent
method at $f \ll f_c$ one finds:
\begin{eqnarray} \label{creep}
V = A~ T ~ {f_c}^{z-1} ~ && (\frac{f}{f_c})^{\frac{z-1-(\gamma/2)}{1-\gamma}}
{}~ \exp( - (1-\gamma) (\frac{\gamma f_c}{f})^{\mu} )
\qquad   \nonumber \\
&&\ \mu=\frac{\gamma}{1-\gamma}
\end{eqnarray}
with $A \sim \sqrt{\gamma (1-\gamma)/2 \pi}$. The exponential
factor holds for any correlator behaving as a power law $\sim x^\gamma$ at
large
distances. The preexponential factor depends on details of the crossover
of the correlator to the logarithmic regime ($x \ll \xi_G)$.
In this creep regime the linear response at $f \to 0$ is
absent and the characteristic barriers which control the
dynamics diverge as $1/f^{\mu}$.

{\it Critical case}
critical behaviour, which generically corresponds to $\gamma=0$, can
equivalently be
achieved by taking $\xi_G = \infty$ in (2).
The $V$ versus $f$ characteristics
becomes a power law at small force. In that case
we have  $K(x) \sim \Delta \log|x|$ at large $x$ and
equation (\ref{disorder}) gives:
\begin{equation} \label{power}
V =  \frac{f^{z-1}}{\Gamma(z-1) T^{2+z}}
\end{equation}
with  $z = 2 + (\Delta/2 T^2)$.
resembling the power law critical behaviour proposed
at the vortex glass transition. This transition separates the
creep dynamics
$V \sim \exp(- const/f^{\mu} )$
in the vortex glass state from the ohmic behaviour $V \propto f$
above the transition. Indeed one finds that the scaling function
$H(y)$ of (\ref{H}) reproduces the correct scaling behaviour
$H(y) \sim y^{1-z}$ in the critical
region  $y \gg 1$ which leads to (\ref{power})
for $T/a^2 \gg f \gg f_c$. Using (\ref{diffcoeff}) we find the critical
behaviour
$D \propto {\xi_G}^{2-z}$.
Since in the vortex system
the voltage is proportional to the vortex velocity and
force is proportional to applied current the obtained behaviour of $D$
reporduces the critical behaviour of the resistivity $\rho$ at the transition,
$\xi_G$ playing the role of the vortex glass critical length.

{\it Ohmic behaviour $\gamma \leq 0$}:
For short-range correlations $\gamma<0$,
$\overline{U(x)^2}$ is finite and one recovers the
Einstein relation (\ref{diffcoeff}) with $D>0$ and
linear response. Let us analyze in detail
the regime of ohmic motion, corresponding to a finite correlation length.
We denote
$\overline{U(x) U(0)}= \Delta~ C(x)$, with $C(0)=1$ and $C(x)$ decays
to zero on a length scale $\xi$.
\begin{equation} \label{short}
\frac{1}{V}  =  \frac{1}{T} \int_0^\infty dz
\exp(- \frac{ f z}{T} + \frac{\Delta}{T^2} (1-C(z) ) )
\end{equation}
Two ohmic regimes, one at small
force with $V \propto e^{-\Delta/T^2} f$ and one at large force
with $V \propto f$, appear. At high temperature these two regimes match
smootly. At low temperature there is a sharp crossover.
A depinning temperature $T_p = \sqrt{\Delta}$ separates
these two different behaviours which we call ''unpinned'' and
''pinned'' respectively.
For $T>>T_p$ the particle motion is mostly unpinned and the smooth
crossover occurs at $f_c \sim T/\xi$.
For $T<<T_p$ a new characteristic force $f_p$ arises which marks
the crossover between $V \propto e^{-\Delta/T^2} f$
for $f<< f_p$ and $V \propto f$ for $f>> f_p$. To estimate $f_p$
we use $C(z)=exp(-(z/\xi)^2)$ which leads to
$f_p \approx (T/\xi) \sqrt{\Delta/T^2} \sim \sqrt{\Delta}/\xi$.
Note that $f_p$ does not depend on temperature and has thus a
limit when $T \to 0$. At $T \to 0$ the curve $V$ versus $f$ becomes
$V = \theta(f-f_p) (f-f_p)$. It is interesting to mention that this
behaviour can be obtained correctly using the perturbation
theory analogous to that used for calculation of
critical currents in type II superconductors \cite{revue_russes} . Disorder
induced corrections to the velocity is given by
$\delta v/v = \Delta \xi v T^{-3} ( \frac{T}{\xi v} - arctan(\frac{T}{\xi
v}))$.
Equating $\delta v/v \approx 1$ we recover the above result
for $f_p$ at low temperature $T<<T_p$ while for $T>>T_p$,
$\delta v/v$ never reaches $1$ and thus $f_p$ does not exist.
This analysis assume a finite mean squared force
$<(\nabla U)^2> \sim {f_p}^2 < \infty $, i.e that the random potential is
smooth.
If we choose instead $C(z)=exp(-|z/\xi|)$ an exact calculation
from (\ref{short}) gives:
$$
V= \frac{T}{\xi \gamma(f\xi/T,1) }  (\frac{\Delta}{2T^2})^{f\xi/T}
exp(-\Delta/2T)
$$
where $\gamma$ is the incomplete Gamma function. In that case
the average force is infinite and at $T=0$ the particle remains pinned
at any applied force.

The above model may apply directly to a string
(directed polymer) of length
$L$ in $d$ dimensions driven by its tip at $z=0$ over
point like impurities.
An example is a flux line in a superconductor
in presence of an external current $J$.
Because of screening at small fields the external Lorentz force
of modulus $f \propto J$ is
exerted only on the end of the line, over a length of order
$\lambda$. Let $F(r)$ be the free energy corresponding
to equilibrium positions $r$ of
the tip at $z=0$ with the other end $z=-L$ fixed at $r=0$
for large $L$. One can view the tip as
a particle driven in the rough potential $F(r)$
with $\overline{(F(r)-F(0))^2} \sim r^{\gamma}$
and \cite{huse_henley,kardar_parisi_zhang} $\gamma= 2 \theta/\zeta$.
Moving the tip by $r$ provokes a reorganization
of the string up to length $z=r^{1/\zeta}$ away from
the tip. It thus leads to correlated barriers and
to the $V$-$f$ characteristics found above.
Since $\zeta \approx 6/(8+d)$
and $\theta=2 \zeta-1$ one has $\gamma \approx (4-d)/3$. Thus
$\gamma=1$ in $d=1+1$, an exact result,
and $\gamma \approx 2/3$ in $d=2+1$. If this picture
is qualitatively correct a flux line driven by its tip in $d=1+1$
would experience a threshold.
Below $f_c$ the motion of the tip would be as $r \sim t^{f/f_c}$
and $V=0$ and $V \sim \exp(-1/f^{2})$ in $d=2+1$.
Of course the true motion is more complicated
and the string might not be assumed to be in equilibrium.

Note that in one dimensional motion the velocity
can be dominated by rare large barriers,
as is the case for the exponent $\mu$ in (\ref{creep}),
whereas diffusion at zero force $f=0$,
given \cite{durrett,bouchaud_correlated} by $x \sim (\log t)^{2/\gamma}$,
is dictated by {\it typical barrier} $E_b \sim x^{\gamma/2}$,
In higher dimensional
space, such as the configurational space of a string,
there are many paths in parallel
from one point to another which may allow to avoid
atypically high barriers. This could effectively
cut the tails of the distribution of barriers,
and the gaussian assumption may not be justified.
This effect can be accounted for within our one dimensional
model by allowing more general distributions
for barriers. Averaging
(\ref{velocity}) over disorder, one can write:
\begin{equation}
\frac{1}{V}  =  \frac{1}{T} \int_0^\infty dz \int dE_b
P(E_b,z)
\exp((E_b - f z)/T )
\end{equation}
where $P(E_b,z)$ is the probability of a
barrier $E_b=U(x+z)-U(x)$ corresponding to separation $z$.
The previous calculation corresponds to a Gaussian
$P(E_b,z) \sim \exp(-{E_b}^2/2 K(z))$.
A more general form is:
\begin{equation}
P(E_b,z)= z^{-\gamma/2} Q(E_b/z^{\gamma/2})
\end{equation}
The behaviour at small $f$ is governed by the large
barriers. Thus we suppose that
$Q(u) \sim \exp(-u^{\delta}/2 \Delta)$ at large $u$.
Then using the saddle point method one obtains:
\begin{equation}
V  \sim \exp(- C T^{-(1+\nu)} f^{-\mu})  \qquad
\mu^{-1}=\frac{2(\delta-1)}{\gamma \delta} - 1
\end{equation}
with $\nu=1/(\delta-1- \frac{\gamma \delta}{2})$ and
$C=\nu^{-1} (2/\gamma \delta)^{1+\nu} (2 \Delta)^{-1/\nu}$.
This is the most general case containing the
Gaussian case for $\delta=2$.
For $\delta \to \infty$, i.e when the distribution
has no tails, the typical barriers
dominate and $\mu=\mu_{typ}=\gamma/(2-\gamma)$.
This calculation shows however that
even for rapidly falling tails, $\delta$ large but finite,
the exponent $\mu$ can be different from
the value $\mu_{typ}$ obtained by considering
typical barriers.

In conclusion, we found stretched exponential velocity versus force
characteristics $V \sim \exp(- const/f^{\mu} )$ in a simple model
of a particle diffusing in a 1D random potential
with correlated barriers. There is a transition between this creep
motion and ohmic motion, characterized by power law
critical behaviour at the transition.
This transition occurs as the correlations in the random potential change
from short range to long range.
We stressed the physical analogies with the creep motion of
driven flux lines in a glassy state, also characterized by correlated
barriers. The transition found here is similar to the vortex
glass transition which can thus be viewed as a transition from the
short range (in the liquid) to the long range correlated behaviour
(in a glassy state) of the effective potential seen by a vortex
configuration. The above results also establish a connection between spin and
vortex glasses.  Our model exhibits both creep behavior specific to vortex
glasses and aging, which is recognized as the essential characteristic of spin
glass dynamics, upon tuning the degree of correlation of random field.

VMV acknowledges support
from Argonne National Laboratory through the U.S.
Department of Energy, BES-Material Sciences,
under contract No. W-31-109-ENG-38.
This research was supported in part by NSF Grant No. PHY89-04035.

{\it Note added:} After submission we received a preprint
by S. Scheidl who also derived Eq. 8. in a different
context.

\end{document}